\PassOptionsToPackage{unicode}{hyperref}
\PassOptionsToPackage{hyphens}{url}
\PassOptionsToPackage{dvipsnames,svgnames,x11names}{xcolor}
\documentclass[
]{article}
\usepackage{amsmath,amssymb}
\usepackage{arxiv}
\usepackage{lmodern}
\usepackage{iftex}
\ifPDFTeX
  \usepackage[T1]{fontenc}
  \usepackage[utf8]{inputenc}
  \usepackage{textcomp} 
\fi
\IfFileExists{upquote.sty}{\usepackage{upquote}}{}
\IfFileExists{microtype.sty}{
  \usepackage[]{microtype}
  \UseMicrotypeSet[protrusion]{basicmath} 
}{}
\makeatletter
\@ifundefined{KOMAClassName}{
  \IfFileExists{parskip.sty}{%
    \usepackage{parskip}
  }{
    \setlength{\parindent}{0pt}
    \setlength{\parskip}{6pt plus 2pt minus 1pt}}
}{
  \KOMAoptions{parskip=half}}
\makeatother
\usepackage{xcolor}
\usepackage{graphicx}
\makeatletter
\def\maxwidth{\ifdim\Gin@nat@width>\linewidth\linewidth\else\Gin@nat@width\fi}
\def\maxheight{\ifdim\Gin@nat@height>\textheight\textheight\else\Gin@nat@height\fi}
\makeatother
\setkeys{Gin}{width=\maxwidth,height=\maxheight,keepaspectratio}
\makeatletter
\def\fps@figure{htbp}
\makeatother
\providecommand{\tightlist}{%
  \setlength{\itemsep}{0pt}\setlength{\parskip}{0pt}}
\NewDocumentCommand\citeproctext{}{}
\NewDocumentCommand\citeproc{mm}{%
  \begingroup\def\citeproctext{#2}\cite{#1}\endgroup}
\makeatletter
 \let\@cite@ofmt\@firstofone
 \def\@biblabel#1{}
 \def\@cite#1#2{{#1\if@tempswa , #2\fi}}
\makeatother
\newlength{\cslhangindent}
\newlength{\csllabelwidth}
\newenvironment{CSLReferences}[2] 
 {\begin{list}{}{%
  \setlength{\itemindent}{0pt}
  \setlength{\leftmargin}{0pt}
  \setlength{\parsep}{0pt}
  \ifodd #1
   \setlength{\leftmargin}{\cslhangindent}
   \setlength{\itemindent}{-1\cslhangindent}
  \fi
  \setlength{\itemsep}{#2\baselineskip}}}
 {\end{list}}
\usepackage{calc}

\def\languageshorthands#1{}

\IfFileExists{bookmark.sty}{\usepackage{bookmark}}{\usepackage{hyperref}}
\IfFileExists{xurl.sty}{\usepackage{xurl}}{} 
\hypersetup{
  pdftitle={Argus: JAX state-space filtering for gravitational wave
detection with a pulsar timing array},
  pdfauthor={Tom Kimpson, Nicholas O'Neill, Patrick M.
Meyers, Andrew Melatos},
  pdflang={en-US},
  colorlinks=true,
  linkcolor={Maroon},
  filecolor={Maroon},
  citecolor={Blue},
  urlcolor={Blue},
  pdfcreator={LaTeX via pandoc}}

\title{Argus: JAX state-space filtering for gravitational wave detection
with a pulsar timing array}
\renewcommand{\headeright}{}
\renewcommand{\shorttitle}{Argus}

\definecolor{c53baa1}{RGB}{83,186,161}
\definecolor{c202826}{RGB}{32,40,38}

\usepackage[affil-it]{authblk}
\author[1,2]{Tom Kimpson}
\author[1,2]{Nicholas O'Neill}
\author[3]{Patrick M. Meyers}
\author[1,2]{Andrew Melatos}

\affil[1]{School of Physics, University of Melbourne, Parkville, VIC
3010, Australia%
    \,\protect\href{https://ror.org/01ej9dk98}{}\,%
  }
\affil[2]{Australian Research Council Centre of Excellence for
Gravitational Wave Discovery (OzGrav),Parkville, VIC 3010, Australia%
    \,\protect\href{https://ror.org/02zp3yd51}{}\,%
  }
\affil[3]{Theoretical Astrophysics Group, California Institute of
Technology, Pasadena, CA 91125, USA%
    \,\protect\href{https://ror.org/05dxps055}{}\,%
  }

\date{13 October 2025}

\begin{document}
\maketitle

\section{Summary}\label{summary}

\texttt{Argus}\footnote{\href{https://github.com/tomkimpson/Argus}{github.com/tomkimpson/Argus}} is a high-performance Python package for detecting and
characterizing nanohertz gravitational waves in pulsar timing array
(PTA) data. The package provides a complete Bayesian inference framework
based on state-space models, using Kalman filtering for efficient
likelihood evaluation. \texttt{Argus} leverages the JAX library
(\citeproc{ref-jax2018github}{Bradbury et al., 2018}) for just-in-time
(JIT) compilation, GPU acceleration, and end-to-end automatic
differentiation, facilitating rapid Bayesian inference with
gradient-based samplers. The state-space approach provides a
computationally efficient alternative to traditional frequency-domain
methods, offering linear scaling with the number of pulse
times-of-arrival, and natural handling of non-stationary processes.

\section{Statement of Need}\label{statement-of-need}

PTAs monitor the precise arrival times of radio pulses from a collection
of millisecond pulsars distributed across the sky. By measuring the
correlated variations in these pulse arrival times PTAs are senstive to
gravitational-waves in a frequency band inaccessible to ground-based
interferometers. The possible discovery of a nanohertz stochastic
gravitational-wave background (GWB) by PTA collaborations
(\citeproc{ref-nanograv2023pta}{NANOGrav Collaboration, 2023})
(\citeproc{ref-EPTA}{EPTA Collaboration et al., 2023})
(\citeproc{ref-ParkesPPTA2023}{Reardon et al., 2023}) through measuring
the spatial correlation of the variations between different pulsars -
the characteristic Hellings-Downs curve
(\citeproc{ref-hellings1983gravitational}{Hellings \& Downs, 1983}) -
represents a landmark achievement in gravitational wave astronomy.

Traditional PTA data-analysis methods operate in the frequency domain.
The various noise processes are treated as Gaussian stationary
processes, characterised by their power spectral densities (PSDs). These
noise sources generally fall into two categories: uncorrelated white
noise and time-correlated red noise. White noise sources include
measurement noise from telescope receivers, while red noise components
include pulsar spin noise (intrinsic to the neutron star) and dispersion
measure (DM) variations (from electron density fluctuations in the
interstellar medium). The GWB signal itself is also modeled as a red
noise process with a characteristic power-law PSD
(\citeproc{ref-Goncharov2021}{Goncharov et al., 2021}), distinguished
from the other noise components by its specific spatial correlation, the
Hellings-Downs correlation. Frequency domain modelling is the foundation
for widely used packages such as ENTERPRISE
(\citeproc{ref-enterprise2020}{Ellis et al., 2020}) and TempoNest
(\citeproc{ref-lentati2013hyper}{Lentati et al., 2014}) and is typically
combined with standard Bayesian inference methods
(\citeproc{ref-2009MNRAS.395.1005V}{van Haasteren et al., 2009}), such
Markov Chain Monte Carlo (MCMC) algorithms for parameter estimation and
model selection.

State-space methods provide a novel, powerful and complementary
framework for PTA data analysis. The approach features a time-domain
version of the Gaussian processes framework, offering an alternative
computational structure to traditional frequency-domain modeling.
Instead of relying on full matrix inversions of the covariance matrix,
state-space methods model the temporal evolution of hidden states (such
as pulsar spin fluctuations and gravitational-wave effects) using Kalman
filtering (\citeproc{ref-kalman1960new}{Kalman, 1960}) for recursive
state estimation. This approach exhibits linear scaling
\(\mathcal{O}(N)\) with the number of observations \(N\). State-space
methods can easily incorporate physical knowledge about how different
stochastic processes evolve over time directly into the model structure,
naturally accommodating non-stationary processes. Additionally, the
method tracks the actual, measured, time-ordered realization of
intrinsic timing noise in each pulsar, rather than averaging over
ensemble realizations, and can readily handle non-Gaussian statistics
(\citeproc{ref-2024arXiv240500058U}{Uhlmann \& Julier, 2024}).

Despite their theoretical advantages, state-space methods have seen
limited adoption in PTA research, partly due to their recency, and
partly due to the lack of accessible, high-performance implementations.
\texttt{Argus} provides a modern, science-ready implementation of
state-space methods for gravitational-wave detection in PTA data. Argus
leverages JAX's just-in-time compilation, automatic differentiation, and
GPU acceleration capabilities to handle the computational demands of
Bayesian inference at PTA scales. Argus consolidates and formalises the
state-space methodology applied in prior work
(\citeproc{ref-kimpson2024a}{Kimpson et al., 2024a},
\citeproc{ref-kimpson2024b}{2024b}, \citeproc{ref-kimpson2025c}{2025}),
transforming proof-of-concept implementations into a tool ready for
scientific analysis.

\section{Relation to Existing Work}\label{relation-to-existing-work}

Frequency-domain PTA packages such as ENTERPRISE
(\citeproc{ref-enterprise2020}{Ellis et al., 2020}) and TempoNest
(\citeproc{ref-lentati2013hyper}{Lentati et al., 2014}) represent timing
noise and GW signals as Gaussian processes specified by power spectra.
In contrast, \texttt{Argus} adopts a time-domain, state-space
formulation in which latent variables describe the pulsar rotational
states and other stochastic processes (e.g., DM variations), and the
likelihood is evaluated via a Kalman filter, with
\(\mathcal{O}\left(N\right)\) complexity
(\citeproc{ref-kalman1960new}{Kalman, 1960}). The Kalman filter tracks
the actual, measured, time-ordered realization of intrinsic, achromatic
timing noise in each pulsar, effectively following the specific random
draw of noise present in the data, which allows the method to separate
and identify GW-induced timing perturbations from this intrinsic noise.
This differs from frequency-domain approaches that characterise timing
noise statistically by fitting a power spectral density, effectively
averaging over an ensemble of admissible noise realizations through a
PSD fit (e.g., Goncharov et al. (\citeproc{ref-Goncharov2021}{2021})).
Prior PTA state-space prototypes established feasibility on mock
datasets (\citeproc{ref-kimpson2024a}{Kimpson et al., 2024a},
\citeproc{ref-kimpson2024b}{2024b}). \texttt{Argus} consolidates this
methodology into a production-ready JAX implementation with JIT/GPU
acceleration and end-to-end automatic differentiation, which enables
gradient-based samplers. As such, the package is complementary to
ENTERPRISE/TempoNest: it offers an independent cross-check with
different numerical/systematic failure modes, while retaining parity in
astrophysical content (white/red noise, DM, and a GWB with
Hellings--Downs correlations). While \texttt{Argus} currently focuses on
the stochastic GWB, the same state-space machinery naturally extends to
deterministic sources such as individual supermassive black hole
binaries (see Future Directions).

\section{Functionality}\label{functionality}

\texttt{Argus} is built on JAX, enabling high-throughput computation on
CPUs/GPUs/TPUs with JIT compilation and end-to-end automatic
differentiation. Its core deliverable is a JAX-jittable log-likelihood
for PTA datasets, making it directly suitable for gradient-based
Bayesian inference.

\textbf{Core functionality}

\begin{itemize}
\tightlist
\item
  \textbf{State-space model construction:} Stochastic processes like
  pulsar-intrinsic red noise (modeled as an Ornstein-Uhlenbeck process)
  and the stochastic GWB are specified in the time domain as linear
  stochastic differential equations (SDEs). The package compiles these
  into a single state-space model, which naturally separates process
  noise (e.g., physical spin wandering and GWB fluctuations) from
  measurement noise (EFAC, EQUAD). Hellings-Downs spatial correlations
  are implemented through the covariance structure that couples the
  stochastic processes across pulsars.

\item \textbf{Kalman filter likelihood evaluation:} The core of the package is
an optimised Kalman filter (\citeproc{ref-kalman1960new}{Kalman, 1960}).
The filter evaluates the likelihood of the time-of-arrival data in the
time domain and achieves a computational complexity that scales linearly
with the number of observations, \(\mathcal{O}(N)\).

\item
  \textbf{Sampler Integration:} The likelihood and its gradients
  (provided by JAX's autodiff capabilities) integrate directly with
  JAX-native samplers such as those in \texttt{numpyro}
  (\citeproc{ref-phan2019composable}{Phan et al., 2019}) or
  \texttt{blackjax}(\citeproc{ref-cabezas2024blackjax}{Cabezas \&
  others, 2024}). This enables the use of efficient gradient-based
  algorithms like Hamiltonian Monte Carlo (HMC).
\item
  \textbf{Standardized Data Input:} \texttt{Argus} ingests pulsar timing
  data through an interface with \texttt{libstempo}
  (\citeproc{ref-vallisneri2020libstempo}{Vallisneri, 2020}), ensuring
  compatibility with standard PTA data formats.
\end{itemize}

\begin{figure}
\centering
\includegraphics{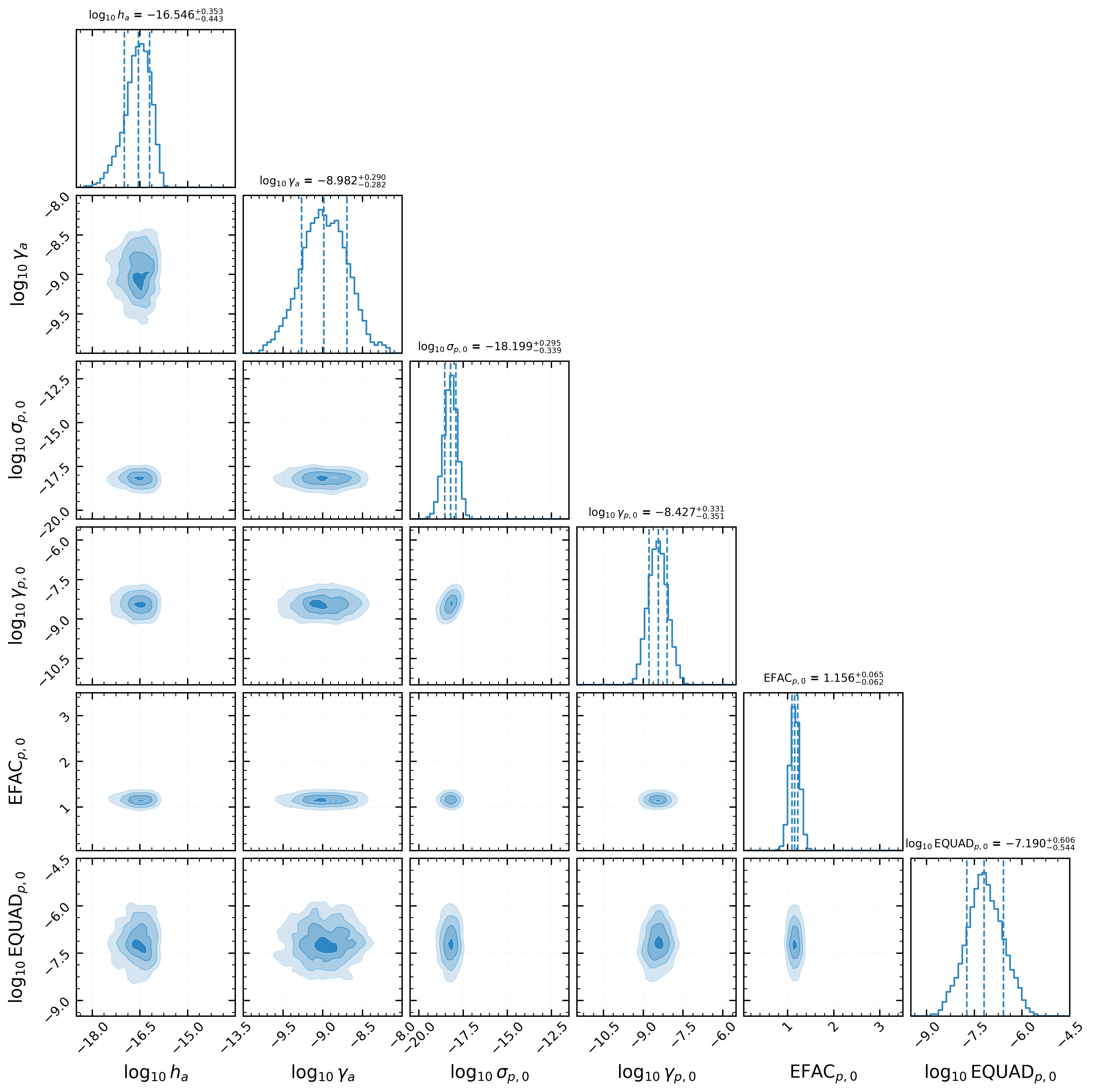}
\caption{Corner plot showing posterior distributions from Bayesian
parameter estimation using Argus on the second IPTA mock data challenge
(\citeproc{ref-2018arXiv181010527H}{Hazboun et al., 2018}). The first
two parameters \(h_{\rm a}\), \(\gamma_{\rm a}\) describe the amplitude
and turnover frequency of the GWB. The middle two parameters
\(\sigma_{\rm p,0}\), \(\gamma_{\rm p,0}\) characterise the red timing
noise for an arbitrary pulsar in the array (indexed by 0). The final two
parameters, EFAC, EQUAD are the standard white measurement noise
parameters for the arbitrary pulsar. The posteriors were obtained using
the No-U-Turn Sampler (NUTS) (\citeproc{ref-hoffman2014no}{Hoffman \&
Gelman, 2014}) from \texttt{numpyro}
(\citeproc{ref-phan2019composable}{Phan et al., 2019}), leveraging
Argus's JAX-native log-likelihood and automatic differentiation for
gradient-based sampling. The unimodal marginalised posteriors
demonstrate the effectiveness of the state-space Kalman filtering
approach for parameter estimation in pulsar timing array
analysis.\label{fig:corner}}
\end{figure}

\section{Future Directions}\label{future-directions}

Planned extensions to \texttt{Argus} include: model selection via Bayes
factors, deterministic signals from continuous gravitational waves,
advanced noise modeling (chromatic noise, non-Gaussian components),
enhanced modularity for custom state-space models, performance
optimization for next-generation PTAs like SKA, and integration with
pipelines such as ENTERPRISE (\citeproc{ref-enterprise2020}{Ellis et
al., 2020}). We release Argus in its current form as it successfully
addresses the core challenge of Bayesian parameter estimation for PTA
analysis using state-space methods, providing a well-tested,
production-ready implementation. We encourage contributions from the PTA
community to help implement these extensions.

\section{Acknowledgements}\label{acknowledgements}

We acknowledge support from the Australian Research Council Centre of
Excellence for Gravitational Wave Discovery (OzGrav) under grant
CE170100004. NJO is the recipient of a Melbourne Research Scholarship.
This work was performed on the OzSTAR national facility at Swinburne
University of Technology. The OzSTAR program receives funding in part
from the Astronomy National Collaborative Research Infrastructure
Strategy (NCRIS) allocation provided by the Australian Government, and
from the Victorian Higher Education State Investment Fund (VHESIF)
provided by the Victorian Government. This work was supported by
software support resources awarded under the Astronomy Data and
Computing Services (ADACS) Merit Allocation Program. ADACS is funded
from the Astronomy National Collaborative Research Infrastructure
Strategy (NCRIS) allocation provided by the Australian Government and
managed by Astronomy Australia Limited (AAL). We thank the International
Pulsar Timing Array collaboration for providing mock data challenges
that aided in the development and validation of this software.

\section*{References}\label{references}
\addcontentsline{toc}{section}{References}

\phantomsection\label{refs}

\end{document}